# Serial coherent diffraction imaging of dynamic samples based on inter-frame constraint

PENGJU SHENG,[1] FUCAI ZHANG[1,*]

[1] *Department of Electrical and Electronic Engineering, Southern University of Science and Technology(SUSTech), Shenzhen 518055, China*
*zhangfc@sustech.edu.cn

We proposed a novel approach to coherent imaging of dynamic samples. The inter-frame similarity of the sample's local structures is found to be a powerful constraint in phasing a sequence of diffraction patterns. We devised a new image reconstruction algorithm that exploits this inter-frame constraint enabled by an adaptive similar region determination approach. We demonstrated the feasibility of this technique in visible light experiments with various real samples, achieving reconstructions of good quality within a few hundred iterations. With a setup as simple as conventional coherent diffraction imaging but with much-improved convergence and robustness to missing data and noise, our method is expected to enrich X-ray imaging techniques and electron microscopy, offering a new tool for dynamics studies.

## 1. INTRODUCTION

Imaging techniques and microscopy play a pivotal role in science and industrial applications. Radiations of short wavelengths like X-rays and high-energy electrons are the primary means for imaging at nanometer or atomic resolutions. Despite steady progress in the development of high-quality optical imaging components for short-wavelength radiations, such as the Fresnel zone plate with sub-ten nanometer outmost zones and Laue mirrors for X-rays, and aberration correctors for high-energy electrons, their performances still fall short of the prevalent demands in materials science and biology. The further enhancement of these components is hampered by the lack of suitable materials and the current limitations of fabrication technology.

In the last 30 years, rapid growth has been seen in lensless coherent diffraction imaging (CDI) techniques [1–8]. As pointed out in [1], once the continuous diffraction pattern of a non-periodic sample is sampled finer than its Nyquist rate, it is possible to reverse the diffraction pattern to obtain the image of the sample. The most used reconstruction algorithms originated from the Gerchberg-Saxton-Fienup algorithms [5,6]. Though those algorithms could work for certain types of samples and data of good quality, they could likely fail for general complex samples.

One recent milestone of CDI is the development of ptychography [9–12]. Ptychography exploits the spatial overlap constraint in the object plane, which greatly improves the convergence of the iterative phasing algorithm and has had significant impacts in many fields [13–18]. In the data acquisition process of ptychography, the overlap in the scan positions leads to considerable redundant information in the acquired diffraction patterns. This redundant information circumvents the twin-image problem in phase retrieval algorithms. It also relaxes the requirements of experimental conditions, such as accurate knowledge of scanning positions [19–21] and illumination coherence [22–24].

For a video recording of the diffraction of an object, considerable redundant information would exist among its frames if the acquisition is faster than the sample's variation rate. Recently, several studies have proposed leveraging the consistency of stationary regions of a sample or employing low-rank priors as novel constraints for CDI, and they obtained promising results [25–31]. X. Tao et al. and Y. H. Lo et al. proposed dividing the sample into stationary and dynamic regions. The algorithm's convergence is enhanced by utilizing the same stationary region across different time frames as a constraint. This strategy increases the complexity of the experiment and requires identifying the stationary region beforehand, thereby limiting its scope of application [26, 29]. G. N. Hinsley et al. utilized variance and hard thresholding to distinguish between time-dependent and time-independent regions within samples, eliminating the need for prior measurements of stationary regions. Still, its applicability is limited, requiring clear boundaries between stationary and varying regions to facilitate effective discrimination by the algorithm [25,28]. N. Vaswani et al. proposed utilizing low rank as a constraint between frames; however, this method requires encoding the exit waves with a binary mask [30, 31].

Here, we propose a novel method called serialCDI, which can rapidly reconstruct a time-varying object from its serially recorded far-field diffraction patterns, with no requirement on prior-measured stationary regions or well-defined boundaries. SerialCDI assumes that in a video recording of far-field diffraction patterns of dynamic samples, a significant portion of the object exit waves would remain structurally similar during the acquisition. We will show that this inter-frame continuity constitutes a powerful constraint in our serialCDI algorithm and is denoted as the inter-frame constraint in this paper. Embodying this constraint can significantly improve the robustness and convergence of the CDI algorithm.

## 2. METHOD

SerialCDI exploits the continuity of the exit wave's local structure among frames of successive recordings. In the framework of serialCDI assumptions, for continuously recording exit waves, the complex wave distribution, $\psi(r; t_n)$, for time instant $t_n$ can be approximated as:

$$\psi(r; t_n) = \psi_s(r) + \psi_e(r; t_n), \quad (1)$$

$r$ is coordinate vector in the object plane; $n = 1,2,3, \ldots, N$; $N$ is the total number of frames; The meaning of Eq. (1) is that a time-varying exit wave can be formed by a shared component $\psi_s(r)$ and a feature component $\psi_e(r; t_n)$. The shared component $\psi_s(r)$ constitutes the inter-frame continuity, while the feature component $\psi_e(r; t_n)$ represent those features dedicated to the $n^{th}$ frame.

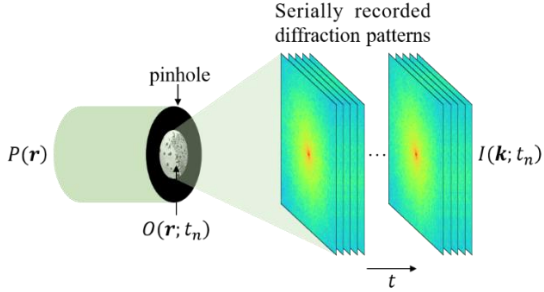

Fig. 1. Schematic layout of experimental setup. The extent of dynamic sample $O(r; t_n)$ is confined by a pinhole; A sequence of far-field diffraction patterns, $I(k; t_n)$, are collected from a time-varying sample.

The core of serialCDI lies in identifying the $\psi_s(r)$ and $\psi_e(r; t_n)$. The revised $\psi_s(r)$, formed by weighted averaging, which carries information from multiple measurements, can generate a more accurate exit wave estimate. The inter-frame constraint is similar to the overlap constraint in ptychography. The difference is that $\psi_s(r)$ refers to a common factor among all exit waves, which is a natural constraint available in many cases, in contrast to ptychography whose overlap constraint is introduced through scanning, limiting its application for dynamic samples. There have been attempts to extend the application of ptychography to dynamic samples [32–34], but these efforts have not yet met expectations. SerialCDI achieves dynamic imaging of samples by utilizing the continuity between frames to construct "overlapping", thereby avoiding scanning. In the following, we introduce a way to realize the extraction of $\psi_s(r)$ and prove its ability to achieve fast convergence and high robustness by simulation and experiments.

Figure 1 is a schematic diagram of the experimental setup. The setup is identical to the traditional CDI, so the proposed method could be conveniently implemented in broad scenarios. A confined probe $P(r)$ is formed on the sample, i.e., by a pinhole. The time-varying object $O(r; t_n)$ is illuminated by $P(r)$; the corresponding exit wave $\psi(r; t_n)$, at various time instants $t_n$, propagates to the detector where their far-field diffraction intensity $I(k; t_n)$ are recorded, where $k$ is the coordinate vector in the detector plane. In addition, a diffraction pattern of the probe, $I_P(k)$, is also acquired.

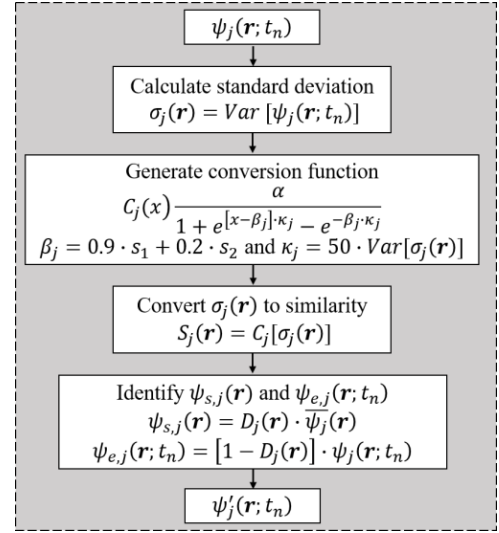

Fig. 2. Procedure for inter-frame constraint. $j$ is the iteration number.

Figure 2 shows the procedure of the inter-frame constraint. In addition to iterating between the real and reciprocal space with the support and modulus constraints applied alternatively, the inter-frame constraint is enforced in the serialCDI as detailed below. After applying the support constraint to all exit waves, the standard deviation map $\sigma_j(r)$ is utilized to measure the degree of variation between exit waves at each pixel. The standard deviation $\sigma_j(r)$ is then converted into the degree of similarity $S_j(r)$ by $S_j(r) = C_j[\sigma_j(r)]$, where

$$C_j(x) = \frac{1}{1 + e^{(x-\beta_j)\kappa_j} - e^{-\beta_j\kappa_j}}. \quad (2)$$

Equation (2) is conceived from the basic form of $1/(1 + e^x)$, which has a flipped $S$ shape. For points in the object plane with smaller standard deviations, the degree of similarity is higher, or vice versa. The term $-e^{-\beta_j\kappa_j}$ is used to ensure $C_j(x)$ reaches 1 when the standard deviation approaches to zero. Due to the continuous nature of the curve $C_j(x)$, serialCDI does not require the region of variation to have a distinct boundary.

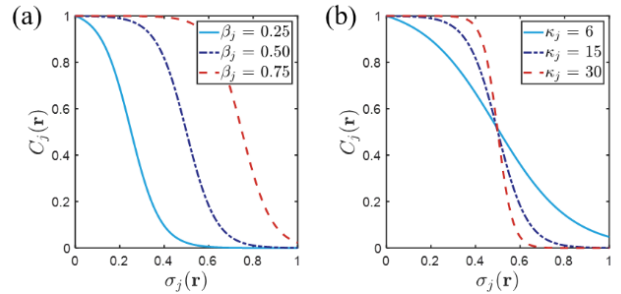

Fig. 3. Effect of $\beta_j$ and $\kappa_j$ on the shape of $C_j(x)$. (a) The shape of $C_j(x)$ when $\kappa_j$ is fixed to 15 and $\beta_j$ takes values of 0.25, 0.5, and 0.75. (b) The shape of the curve, $C_j(x)$, when $\beta_j = 0.5$ and $\kappa_j$ takes values of 6, 15, and 30.

The values of $\beta_j$ and $\kappa_j$ are updated as the iteration progresses. To have a better robustness during iterations, $\beta_j$ is calculated as $\beta_j = 0.9 \cdot s_1 + 0.2 \cdot s_2$ and $\kappa_j$ is set to equal 50 times the standard deviation of $\sigma_j(r)$. Here $s_1$ and $s_2$ are the average corresponding to the lowest 10% and highest 10% elements of $\sigma_j(r)$. Using averages instead of the highest and lowest values can achieve a more stable convergence because the latter tends to vary wildly between iterations.

As depicted in Fig. 3, the value of $\beta_j$ governs the width of $C_j(x)$, while $\kappa_j$ regulates its slope. The distribution of $S_j(r)$ with values close to one identify regions of higher similarity, and multiply it with the inter-frame average of $\psi_j(r;t_n)$ will provide an estimate for $\psi_{s,j}(r)$,

$$\psi_{s,j}(r) = S_j(r) \cdot \overline{\psi_j}(r). \quad (3)$$

Multiplying $1 - S_j(r)$ with the exit waves yields the term $\psi_{e,j}(r;t_n)$, i.e., $\psi_{e,j}(r;t_n) = [1 - S_j(r)]\psi_j(r;t_n)$. Then combining the two terms yields a revised exit wave guess,

$$\psi'_j(r;t_n) = \psi_{e,j}(r;t_n) + \sqrt{\frac{\sum|S_j(r)\psi_j(r;t_n)|^2}{\sum|\psi_{s,j}(r)|^2}} \psi_{s,j}(r). \quad (4)$$

Here, a factor is imposed on $\psi_{s,j}(r)$ to ensure energy conservation. Taking the Fourier transform to the revised exit wave guesses and applying the modulus constraint, an updated diffracted wave is obtained. Applying the inverse Fourier transform to the updated wave produces new exit waves, and the iteration is continued. After repeating the aforementioned steps hundreds of times, the iteration can usually reach a preliminary convergence. At this stage, the initial guesses for the object and probes are generated. The initial guess for the probe is the time-average of $\psi_j(r;t_n)$,

$$P_j(r) = \overline{\psi_j}(r). \quad (5)$$

The initial guess for the object is:

$$O_j(r;t_n) = \frac{\psi_j(r;t_n)}{P_j(r) + \epsilon}, \quad (6)$$

where $\epsilon$ is a small constant preventing a division by zero. After generating the initial guesses for $O_j(r;t_n)$ and $P_j(r)$, the following updating like the ePIE engine [11] in ptychography is run to separate out the $O_j(r;t_n)$ and $P_j(r)$ from the exit-wave,

$$O_{j+1}(r;t_n) = O_j(r;t_n) + \alpha_1 \frac{P_j^*(r)}{|P_j^*(r)|^2_{max}} \cdot \Delta\psi_j, \quad (7)$$

$$P_{j+1}(r) = P_j(r) + \alpha_2 \frac{O_j^*(r;t_n)}{|O_j^*(r;t_n)|^2_{max}} \cdot \Delta\psi_j, \quad (8)$$

where $\Delta\psi_j = [\psi_j''(r;t_n) - \psi_j(r;t_n)]$, $\psi_j''(r;t_n)$ is the exit wave guess after applying modulus constraint and $\psi_j(r;t_n)$ is the exit wave guess before applying inter-frame constraint. The constants of $\alpha_1$ and $\alpha_2$ alter the rate of updating. In addition to being updated through the ePIE engine, $P_j(r)$ is also confined to the support constraint and the modulus constraint. Then multiply the $P_{j+1}(r)$ and $O_{j+1}(r;t_n)$ together to obtain $\psi_{j+1}(r;t_n)$ for the next iteration and repeat the above steps until a predefined stop criteria is met. It is worth noting that we here adopt the difference map updating formula in the modulus constraint [8]. The error reduction (ER) [6] algorithm is utilized at the final stage of reconstruction to seek the global minimum.

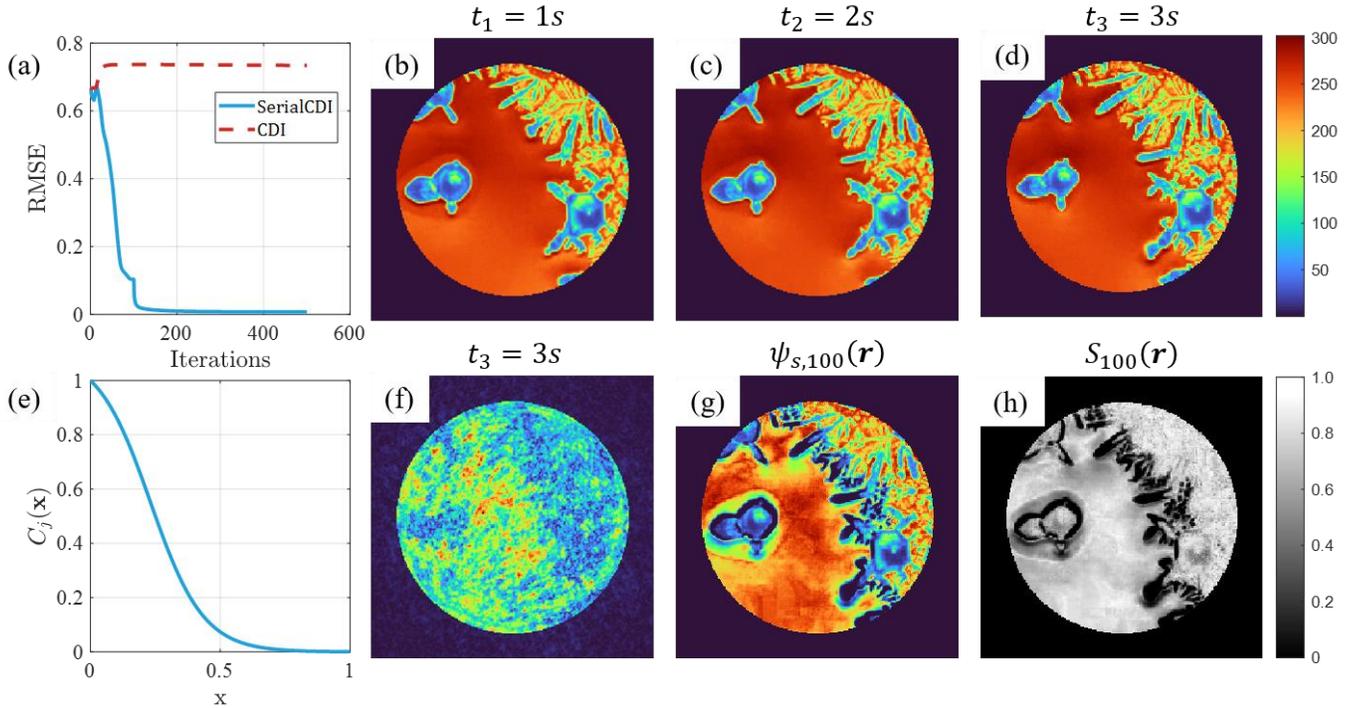

Fig. 4. Simulation of serialCDI and conventional CDI of the crystal growth process. (a) The average RMSE of all exit waves was used to monitor the convergence of the serialCDI (blue continuous line) and conventional CDI (red dash line). (b)-(d) SerialCDI reconstruction results for $t_1 = 1s$, $t_2 = 2s$ and $t_3 = 3s$ respectively. (e) Conversion function $C_j(x)$ at the 100th iteration. (f) CDI reconstruction result for $t_3 = 3s$. (g) Shared component $\psi_{s,100}(r)$ at the 100th iteration. (h) The degree of similarity calculated by $C_j[\sigma_j(r)]$ at the 100th iteration.

## 3. SIMULATION

To validate the reliability of serialCDI in reconstructing dynamic samples, we conducted numerical simulations to image the growth of a crystal sample. A set of 30 far-field diffraction patterns with dimensions of 512×512 pixels were simulated with a detector pixel size of 6.5 μm and a frame rate of 8. Poisson noise was added to the diffraction intensities. The sample was confined within a pinhole with a diameter of 2.3 mm and illuminated by a planar light. The average illumination flux was set to $1.5 \times 10^9$ photons. It is worth noting that in the simulation, only the exit waves at the sample plane are reconstructed for comparison. The invariant normalized error metric (RMSE) proposed in [35] was used to monitor the convergence.

Figure 4 (a) shows the convergence curves of traditional CDI and serialCDI. As one can see, serialCDI exhibits a significantly faster convergence, reaching an RMSE below 0.01 within 200 iterations. The amplitude of the reconstructed amplitudes and phases using the serialCDI algorithm are shown for selected frames at $t_1 = 1s$, $t_2 = 2s$ and $t_3 = 3s$ in Fig. 4(b)-(d) respectively. The result of the traditional CDI at $t_3 = 3s$ are presented in Fig. 4(f), which exhibits noticeable artifacts with little sample features. After 100 iterations, the inter-frame constraints and difference map updating formula were disabled, and the ER algorithm was employed to seek lower RMSE.

Figure 4(g) and (h) show the shared component $\psi_{s,j}(r)$ and similarity $S_j(r)$ at 100th iteration respectively. As can be seen from Fig. 4(b)-(d) and supplementary video 0, areas with large changes in the sample have lower similarity $S_j(r)$, as indicated by the dark region in Fig. 4(h). A more detailed comparison video of reconstructed frames can be found in supplementary video 0.

Similarly, the shared component obtained at corresponding positions is also minimal. This simulation demonstrates that serialCDI can effectively extract shared components and accelerate algorithm convergence via the inter-frame constraint.

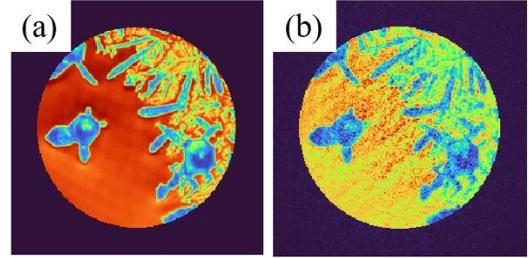

Fig. 5. Effect of $\kappa_j$ on the reconstruction result. (a)-(b) Reconstructions result for dynamic updating $\kappa_j$ and fixing $\kappa_j$ as 10000 respectively.

Figure 5(b) shows the reconstruction result when $\kappa_j$ is set to a fixed value of 10000, which corresponds to a hard-threshold conversion function $C_j(x)$. In this case, , only the shared component with a standard deviation below the threshold will be extracted, and the shared component is generated through direct averaging. If the shared component exhibits minimal energy fluctuations, these fluctuations will be averaged out directly, leading to inaccurate results. On the contrary, a soft-threshold $C_j(x)$ yields a more accurate shared component through weighted averaging. Moreover, updating $\kappa_j$ during iteration expedites the convergence and robustness, as illustrated in Fig. 5 (a).

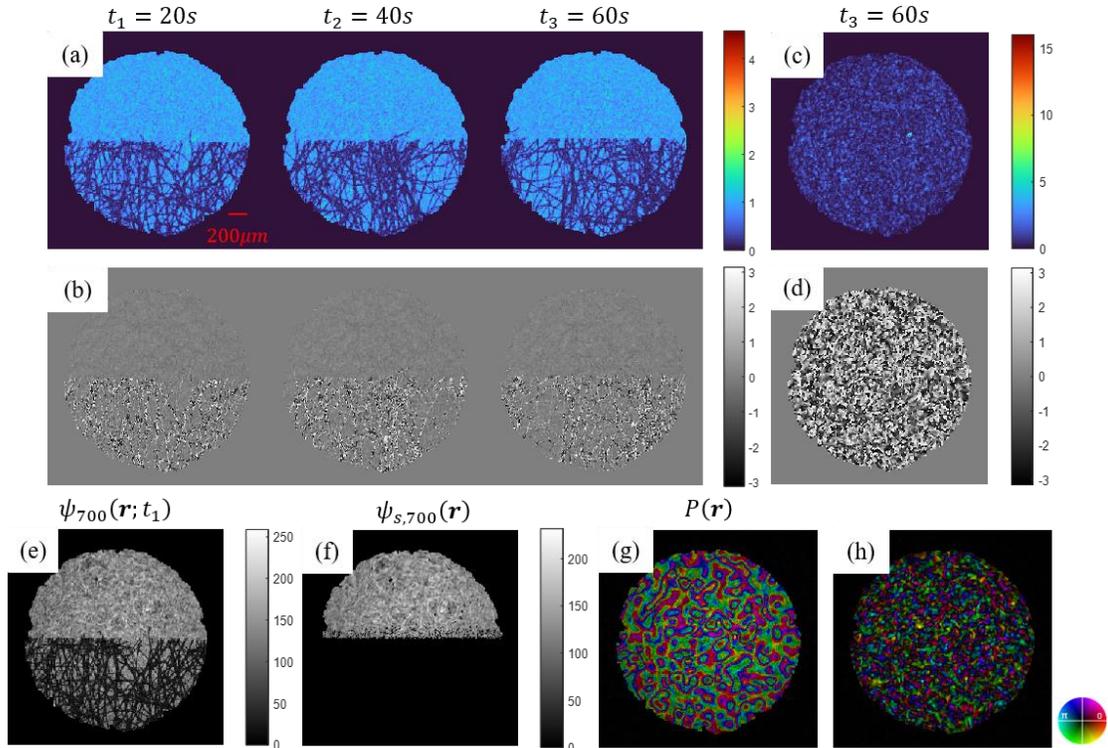

Fig.6. Experimental results of the serialCDI with a tissue sample. (a) Reconstructed amplitude at $t_1 = 20s$, $t_2 = 40s$ and $t_3 = 60s$ using the serialCDI algorithm, respectively. (b) The corresponding phase for (a). (c) and (d) The reconstructed result of traditional CDI at $t_3 = 60s$. (e) The amplitude of the reconstructed exit wave at $t_1 = 20s$ after 700 iterations. (f) Amplitude of shared component $\psi_{s,700}(r)$. (g) Probe reconstructed by serialCDI. (h) Probe reconstructed by traditional CDI. Scale bar: 200μm.

## 4. EXPERIMENT

As a demonstration of the principle, we designed two experiments to showcase the imaging capability of serialCDI for dynamic samples. In the first experiment, a tissue was laterally moved to simulate a dynamic sample. The sample was illuminated with a structured light with a wavelength of 520 nm and a diameter of 2.0 mm and the structured illumination was generated by a randomly distributed modulator placed in front of the pinhole. The downstream of the sample was a lens with a focal length of 75 mm to simulate the far-field recording condition. At the back focal plane of the lens was a detector with a pixel size of 6.5 $\mu m$, 60 diffraction patterns of the sample, and a diffraction pattern of the probe, all with 800×800 pixels, were collected. The frame rate of the detector was set to 1 fps to simulate the case where the sample varies much faster than the camera acquisition speed, and the exposure time was 0.6 ms. The serialCDI and traditional CDI reconstruction results after 2000 iterations are shown in Fig. 6.

The reconstruction results in Fig.6 (a) and (b) showed the amplitudes and phases of the sample at $t_1 = 20s$, $t_2 = 40s$ and $t_3 = 60s$ respectively. It is evident from Fig.6 (c) and (d) that traditional CDI ceases to converge when dealing with complex samples. SerialCDI provides a significantly improved image quality compared to the traditional CDI, both in terms of the reconstruction results of objects and probe (Fig,6 (g) and (h)). Figure 6(e) reveals that at the 700th iteration, the exit waves have been well reconstructed, demonstrating the effectiveness of the inter-frame constraint in achieving fast convergence. As indicated by Fig. 6(f), the lower half, occupied by the rapidly changing sample exhibits low energy, this shows that serialCDI can identify shared component $\psi_s(\mathbf{r})$ accurately.

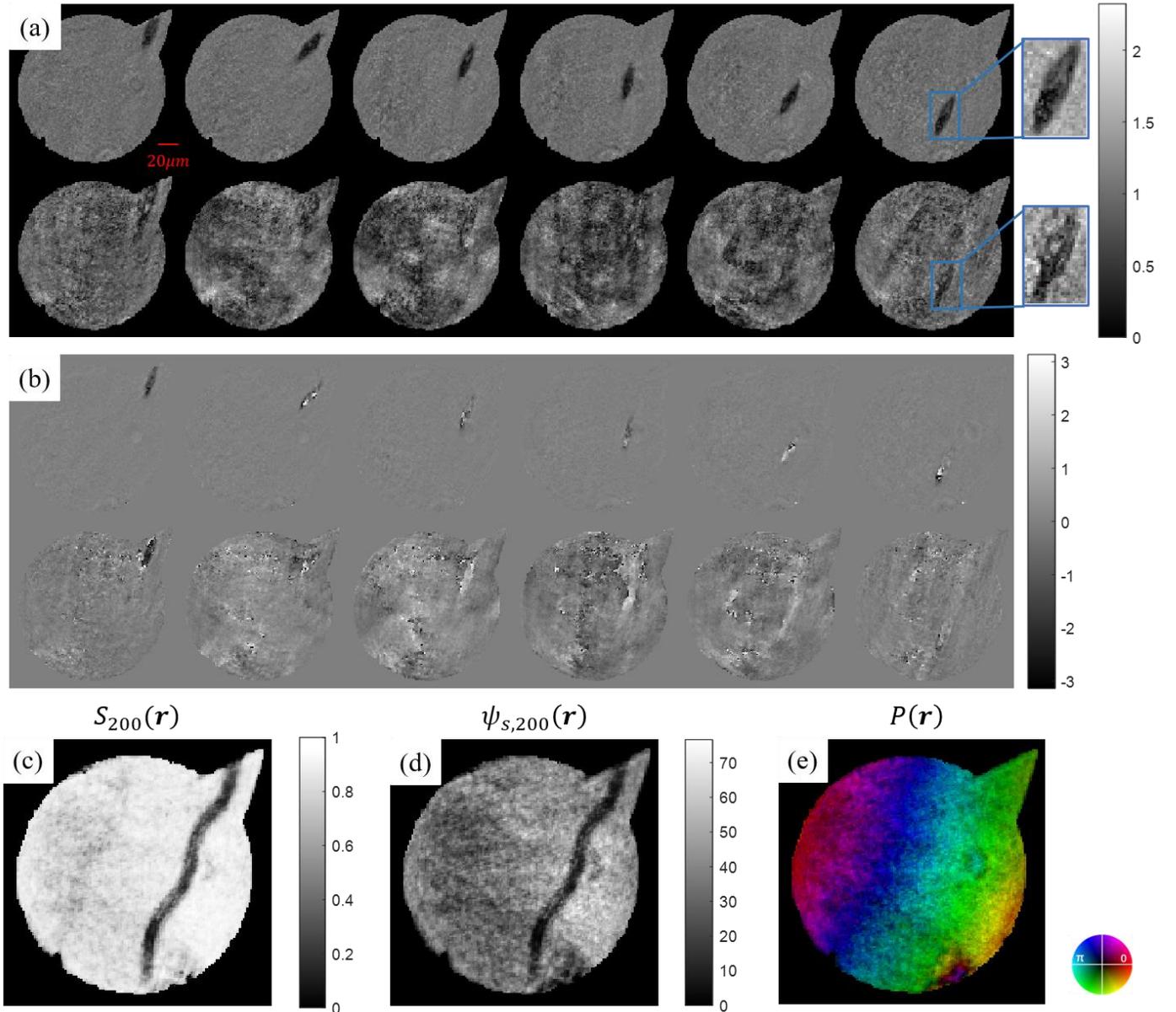

Fig. 7. Experimental results of the serialCDI with a biological sample. (a) The amplitude distributions reconstructed by serialCDI (first row) and CDI (second row) with a time interval of 0.2s. (b) The corresponding phase of (a). (c) Degree of similarity at the 200th iteration. (d) Shared component $\psi_s(\mathbf{r})$. (e) Probe reconstructed by serialCDI. The phase change was caused by inclined illumination. Scale bar: 20μm.

In the presented results, the shared component $\psi_s(\boldsymbol{r})$ is calculated from all frames. In our simulations, three frames would be sufficient to determine $\psi_s(\boldsymbol{r})$. In case of data collection of a long sequence of frames, for which significant changes of sample structure could occur, a running time window can be used to just select a fixed number of frames.

The introduction of unknown structured illumination increased the complexity of the exit waves; however, it can effectively accelerate algorithm convergence by providing a strong inter-frame constraint and improving the signal-to-noise ratio of the diffraction patterns. As shown in supplementary video 1, the energy of the diffraction pattern spread out from the center, resulting in an increased signal-to-noise ratio in the high-frequency region. Therefore, when the rate of change of the sample considerably exceeds the camera acquisition speed, resulting in little similarity information between consecutive frames, the inter-frame constraint can still be kept effective by introducing structured illumination in the object plane. The ability of serialCDI to observe dynamic samples is only limited by the acquisition speed of the camera, while dynamic blurring can be eliminated by short exposure times or short pulse duration. SerialCDI shall be well suitable for x-ray free electron lasers (XFELs), taking advantage of its short pulse duration, and high brightness.

In the second experiment, a planar wave with a wavelength of 518.6nm was used to illuminate swimming euglenids with a typical size from 10 $\mu m$ to 40 $\mu m$. A 10-times magnification optics was placed downstream of the sample to facilitate the observation of finer structures of the euglenids. The transmitting wave through the sample, after magnification, was directed onto a pinhole. This light then passed through a Fourier lens and at its back focal plane, diffraction patterns were captured by a 16-bit camera with 512×512 pixels, each 6.5 $\mu m$ wide, at a rate of 100 frames per second and an exposure time of 5.43 ms. The exposure time was pre-calculated to ensure that the translation of the euglenids remained within one pixel throughout the exposure, avoiding motion blur. A total of 150 diffraction patterns of the sample and one diffraction pattern of the probe were taken. The serialCDI and traditional CDI reconstruction results after 500 iterations are shown in Fig. 7. The experimental optical path differs from the first experimental setup mainly in the inclusion of the magnification system.

As shown in supplementary video 2, by comparing the reconstructed results of the two methods, it can be observed that in the results of CDI, only the amplitude of high-contrast euglenid is visible, while in the results of serialCDI, two euglenids can be seen. This is largely also due to the much uniform background in serialCDI reconstruction. The ability of serialCDI to separate illumination and object functions can be seen more clearly in the phase images in Fig. 7(b).

As can be seen from the similarity map in Fig. 7(c), serialCDI effectively identified low-contrast euglenid from the background during the iteration process. This demonstrates the importance of dynamically updating $\kappa_j$, which helps prevent information loss as would occur when direct averaging was used.

For weakly refracting biological samples as in the second experiment, the primary energy is concentrated in the low-frequency region of the diffraction patterns. To record high-angle dim signals for better resolution, we chose to let the central ~50 pixels per diffraction pattern be overexposed, as highlighted in red in Fig. 8(a). The missing data area was artificially enlarged to test the limit to missing data. Fig. 8(b) highlights the regions artificially masked out, which constituted ~960 pixels (c.a., 28 speckles). Though the image quality degradation can be seen in Fig. 8(c)-(d), the euglenid can still be well identified, demonstrating the increased robustness over the traditional CDI. The results of the reconstruction under three conditions (traditional CDI, serialCDI and data missing serialCDI) can be seen in supplementary video 2.

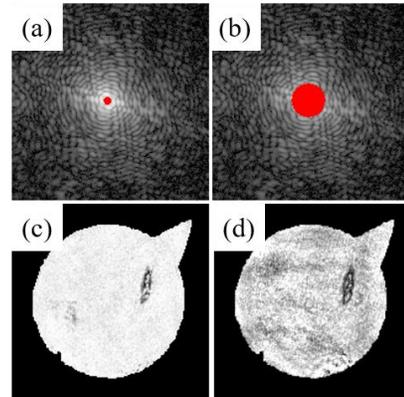

Fig. 8. Robustness of serialCDI to data missing. The diffraction patterns are shown in the log scale. (a) Red pixels indicate overexposed pixels for the result in Fig. 7. (b) Artificially set overexposed pixels with enlarged area, ~960 pixels. (c) Reconstructed amplitude from data (a). (d) Reconstructed amplitude from data (b).

## 5. CONCLUSION

To conclude, we propose an algorithm for imaging dynamical samples called serialCDI, which exhibits rapid convergence and high robustness to missing data in reconstructing experiments data of real samples. We also have demonstrated that the rate of observing dynamic samples depends solely on the camera's acquisition rate. Moreover, short exposure times can mitigate dynamic blur induced by sample vibration.

SerialCDI holds high potential for further improvements. Currently, serialCDI requires the sample exit wave to have a clear boundary. The shrink-wrap algorithm can mitigate this requirement [4], but another improvement is still needed to make serialCDI applicable to generally extended samples. By combining multi-wavelength light sources [36,37] with serialCDI, the imaging quality of current multi-wavelength CDI may be improved, facilitating dynamic imaging with chemically resolved capability. Like ptychography, the serialCDI dataset has a high degree of redundancy. Hence, there is a potential for implementing serialCDI with broadband light sources by adopting mixed states[24], monochromatization [38], polyCDI [39] or deconvolution approaches [40], thereby enhancing its suitability for transmission electron microscopy (TEM) or XFELs, for which partial coherence has been a principal limiting factor affecting imaging resolution [39,41].

The proposed serialCDI method has a simple setup and does not require imaging optics. Therefore, it is applicable across the spectrum of light and particle radiations. A condenser lens with an upstream aperture can form the confined probe, leaving a large working distance around the sample. Thus, serialCDI would be highly suitable for in-situ observation of dynamic samples within their native environment.

**Funding**. National Natural Science Foundation of China (12074167); Shenzhen Science and Technology Innovation

Program (KQTD20170810110313773).

**Disclosures**. Some of the methods described in this paper are the subject of pending patents of the Southern University of Science and Technology.## References

1. J. Miao, T. Ishikawa, I. K. Robinson, and M. M. Murnane, "Beyond crystallography: Diffractive imaging using coherent x-ray light sources," Science **348**, 530–535 (2015).
2. J. Miao, "Extending the methodology of X-ray crystallography to allow X-ray microscopy without X-ray optics," in AIP Conference Proceedings (AIP, 2000), Vol. 507, pp. 581–585.
3. S. Eisebitt, J. Lüning, W. F. Schlotter, M. Lörgen, O. Hellwig, W. Eberhardt, and J. Stöhr, "Lensless imaging of magnetic nanostructures by X-ray spectro-holography," Nature **432**, 885–888 (2004).
4. S. Marchesini, H. He, H. N. Chapman, S. P. Hau-Riege, A. Noy, M. R. Howells, U. Weierstall, and J. C. H. Spence, "X-ray image reconstruction from a diffraction pattern alone," Phys. Rev. B **68**, 140101 (2003).
5. J. R. Fienup, "Phase retrieval algorithms: a comparison," Appl. Opt. **21**, 2758 (1982).
6. J. R. Fienup, "Reconstruction of an object from the modulus of its Fourier transform," Opt. Lett. **3**, 27 (1978).
7. J. S. Wu, U. Weierstall, J. C. H. Spence, and C. T. Koch, "Iterative phase retrieval without support," Opt. Lett. **29**, 2737 (2004).
8. V. Elser, "Phase retrieval by iterated projections," J. Opt. Soc. Am. A **20**, 40 (2003).
9. H. M. L. Faulkner and J. M. Rodenburg, "Error tolerance of an iterative phase retrieval algorithm for moveable illumination microscopy," Ultramicroscopy **103**, 153–164 (2005).
10. H. M. L. Faulkner and J. M. Rodenburg, "Movable Aperture Lensless Transmission Microscopy: A Novel Phase Retrieval Algorithm," Phys. Rev. Lett. **93**, 023903 (2004).
11. F. Hüe, J. M. Rodenburg, A. M. Maiden, and P. A. Midgley, "Extended ptychography in the transmission electron microscope: Possibilities and limitations," Ultramicroscopy **111**, 1117–1123 (2011).
12. A. M. Maiden and J. M. Rodenburg, "An improved ptychographical phase retrieval algorithm for diffractive imaging," Ultramicroscopy **109**, 1256–1262 (2009).
13. Pan X.-Y., Bi X.-X., Dong Z., Geng Z., Xu H., Zhang Y., Dong Y.-H., Zhang C.-L., Beijing Synchrotron Radiation Facility, Institute of High Energy Physics, Chinese Academy of Sciences, Beijing 100049, China, School of Nuclear Science and Technology, University of Chinese Academy of Sciences, Beijing 100049, China, and Spallation Neutron Source Science Center, Institute of High Energy Physics, Chinese Academy of Sciences, Dongguan 523808, China, "Review of development for ptychography algorithm," Acta Phys. Sin. **72**, 054202 (2023).
14. J. Marrison, L. Räty, P. Marriott, and P. O'Toole, "Ptychography – a label free, high-contrast imaging technique for live cells using quantitative phase information," Sci Rep **3**, 2369 (2013).
15. J. M. Rodenburg, A. C. Hurst, A. G. Cullis, B. R. Dobson, F. Pfeiffer, O. Bunk, C. David, K. Jefimovs, and I. Johnson, "Hard-X-Ray Lensless Imaging of Extended Objects," Phys. Rev. Lett. **98**, 034801 (2007).
16. G. R. Brady, M. Guizar-Sicairos, and J. R. Fienup, "Optical wavefront measurement using phase retrieval with transverse translation diversity," Opt. Express **17**, 624 (2009).
17. K. Giewekemeyer, P. Thibault, S. Kalbfleisch, A. Beerlink, C. M. Kewish, M. Dierolf, F. Pfeiffer, and T. Salditt, "Quantitative biological imaging by ptychographic x-ray diffraction microscopy," Proc. Natl. Acad. Sci. U.S.A. **107**, 529–534 (2010).
18. Z. Chen, M. Odstrcil, Y. Jiang, Y. Han, M. Chiu, L. Li, and D. Muller, "Mixed-state electron ptychography enables sub-angstrom resolution imaging with picometer precision at low dose," Nat Commun **11** (2020)
19. M. Beckers, T. Senkbeil, T. Gorniak, K. Giewekemeyer, T. Salditt, and A. Rosenhahn, "Drift correction in ptychographic diffractive imaging," Ultramicroscopy **126**, 44–47 (2013).
20. A. M. Maiden, M. J. Humphry, M. C. Sarahan, B. Kraus, and J. M. Rodenburg, "An annealing algorithm to correct positioning errors in ptychography," Ultramicroscopy **120**, 64–72 (2012).
21. F. Zhang, I. Peterson, J. Vila-Comamala, A. Diaz, F. Berenguer, R. Bean, B. Chen, A. Menzel, I. K. Robinson, and J. M. Rodenburg, "Translation position determination in ptychographic coherent diffraction imaging," Opt. Express **21**, 13592 (2013).
22. D. J. Batey, D. Claus, and J. M. Rodenburg, "Information multiplexing in ptychography," Ultramicroscopy **138**, 13–21 (2014).
23. R. Karl, C. Bevis, R. Lopez-Rios, J. Reichanadter, D. Gardner, C. Porter, E. Shanblatt, M. Tanksalvala, G. F. Mancini, M. Murnane, H. Kapteyn, and D. Adams, "Spatial, spectral, and polarization multiplexed ptychography," Opt. Express **23**, 30250 (2015).
24. P. Thibault and A. Menzel, "Reconstructing state mixtures from diffraction measurements," Nature **494**, 68–71 (2013).
25. G. Hinsley, G. van Riessen, and C. M. Kewish, "Toward dynamic ptychography using a spatiotemporal overlap constraint," Acta Crystallographica Section A Foundations and Advances (2021): n. pag.
26. X. Tao, Z. Xu, H. Liu, C. Wang, Z. Xing, Y. Wang, and R. Tai, "Spatially correlated coherent diffractive imaging method," Appl. Opt. **57**, 6527 (2018).
27. Z. Yuan and H. Wang, "Phase retrieval with background information," Inverse Problems **35**, 054003 (2019).
28. G. N. Hinsley, C. M. Kewish, and G. A. Van Riessen, "Dynamic coherent diffractive imaging using unsupervised identification of spatiotemporal constraints," Opt. Express **28**, 36862 (2020).
29. Y. H. Lo, L. Zhao, M. Gallagher-Jones, A. Rana, J. J. Lodico, W. Xiao, B. C. Regan, and J. Miao, "In situ coherent diffractive imaging," Nat Commun **9**, 1826 (2018).
30. S. Nayer, P. Narayanamurthy, and N. Vaswani, "Provable Low Rank Phase Retrieval," IEEE Transactions on Information Theory **66**, 5875-5903 (2020).
31. N. Vaswani, S. Nayer, and Y. C. Eldar, "Low Rank Phase Retrieval," IEEE Transactions on Signal Processing **65**, 4059-4074 (2017).
32. J. N. Clark, X. Huang, R. J. Harder, and I. K. Robinson, "Dynamic Imaging Using Ptychography," Phys. Rev. Lett. **112**, 113901 (2014).
33. X. Huang, K. Lauer, J. N. Clark, W. Xu, E. Nazaretski, R. Harder, I. K. Robinson, and Y. S. Chu, "Fly-scan ptychography," Sci Rep **5**, 9074 (2015).
34. P. Sidorenko and O. Cohen, "Single-shot ptychography," Optica **3**, 9 (2016).
35. J. R. Fienup, "Reconstruction of a complex-valued object from the modulus of its Fourier transform using a support constraint," J. Opt. Soc. Am. A **4**, 118 (1987).
36. S. Petrakis, A. Skoulakis, Y. Orphanos, A. Grigoriadis, G. Andrianaki, D. Louloudakis, N. Kortsalioudakis, A. Tsapras, C. Balas, D. Zouridis, E. Pachos, M. Bakarezos, V. Dimitriou, M. Tatarakis, E. P. Benis, and N. A. Papadogiannis, "Coherent XUV Multispectral Diffraction Imaging in the Microscale," Applied Science (2022): n. pag.
37. E. Malm, E. Fohtung, A. Mikkelsen. "Multi-wavelength phase retrieval for coherent diffractive imaging." Optics letters **46** 1 (2020): 13-16.
38. J. Huijts, S. Fernandez, D. Gauthier, M. Kholodtsova, A. Maghraoui, K. Medjoubi, A. Somogyi, W. Boutu, and H. Merdji, "Broadband coherent diffractive imaging," Nat. Photonics **14**, 618–622 (2020).
39. B. Abbey, L. Whitehead, H. Quiney, D. Vine, G. Cadenazzi, C. Henderson, K. Nugent, E. Balaur, C. T. Putkunz, A. Peele, Garth J. Williams, I. McNulty, "Lensless imaging using broadband X-ray sources," NATURE PHOTONICS **5**, (2011).
40. H. Lin, W. Xu, J. Zhao, and F. Zhang, "Broadband coherent modulation imaging with no knowledge of the illumination spectrum distribution," Opt. Lett. **48**(15), 3977 (2023).